\def\erg{{\rm\thinspace erg}}

\def\keV{{\rm\thinspace keV}}

\def\Msun{\hbox{$\rm\thinspace M_{\odot}$}}

\def\s{{\rm\thinspace s}}

\def\ergps{\hbox{$\erg\s^{-1}\,$}}

\def\Lx{\hbox{$\thinspace L_\mathrm{X}$}}
\def\Lkin{\hbox{$\thinspace L_\mathrm{kin}$}}
\def\Lnuc{\hbox{$\thinspace L_\mathrm{nuc}$}}
\def\LEdd{\hbox{$\thinspace L_\mathrm{Edd}$}}
\def\Poutflow{\hbox{$\thinspace P_\mathrm{outflow}$}}
\def\Pradiation{\hbox{$\thinspace P_\mathrm{radiation}$}}
\def\Loutflow{\hbox{$\thinspace L_\mathrm{outflow}$}}
\def\Lcluster{\hbox{$\thinspace L_\mathrm{cluster}$}}
\def\Lnucleus{\hbox{$\thinspace L_\mathrm{nucleus}$}}

\def\apj{ApJ}
\def\mnras{MNRAS}

\documentclass{iaus}
\usepackage{graphicx}

\title[JD 11.~~Radiatively-inefficient cluster cores] 
{Highly-luminous cool core clusters of galaxies: mechanically-driven or radiatively-driven AGN?}

\author[Julie Hlavacek-Larrondo \& Andy Fabian]   
{Julie Hlavacek-Larrondo$^{1\star}$
 \and Andy Fabian$^1$}

\affiliation{$^1$Institute of Astronomy, University of Cambridge,\\ Madingley Road, Cambridge CB3 0HA \\ email: {\tt juliehl@ast.cam.ac.uk} \\[\affilskip]}

\pubyear{2011}
\volume{xxx}  
\pagerange{??-??}
\jname{Tracing the Ancestry of Galaxies (on the Land of our Ancestors)}
\editors{A.C. Editor, B.D. Editor \& C.E. Editor, eds.}
\begin{document}

\maketitle

\begin{abstract}
Cool core clusters of galaxies require strong feedback from their central AGN to offset cooling. We present a study of strong cool core, highly-luminous (most with $\Lx\ge~10^{45}~\ergps$), clusters of galaxies in which the mean central AGN jet power must be very high yet no central point X-ray source is detected. Using the unique spatial resolution of $Chandra$, a sample of 13 clusters is analysed, including A1835, A2204, and one of the most massive cool core clusters, RXCJ1504.1-0248. All of the central galaxies host a radio source, indicating an active nucleus, and no obvious X-ray point source. For all clusters in the sample, the nucleus has an X-ray bolometric luminosity below 2 per cent of that of the entire cluster. We investigate how these clusters can have such strong X-ray luminosities, short radiative cooling-times of the inner intracluster gas requiring strong energy feedback to counterbalance that cooling, and yet have such radiatively-inefficient cores with, on average, $\Lkin$/$\Lnuc$ exceeding 200. Explanations of this puzzle carry significant implications for the origin and operation of jets, as well as on establishing the importance of kinetic feedback for the evolution of galaxies and their surrounding medium.
\end{abstract}

\firstsection 
\section{Introduction}
Clusters of galaxies with steeply rising X-ray surface brightness profiles are known as cool core clusters, and require strong feedback from their central AGN to offset cooling of the intracluster medium (ICM). 

\begin{figure}[t]
\begin{center}
 \includegraphics[width=3.in]{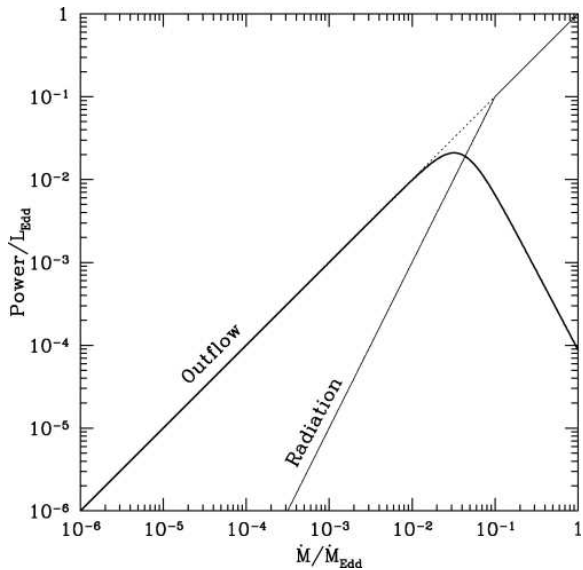} 
 \caption{Sketch of black hole energy release as a function of mass accretion (\cite[Churazov et al. 2005]{Churazov et al. 2005}). The energy in radiation dominates at high accretion rates. If a black hole is releasing $10^{45}\ergps$, then for $M_\mathrm{BH}\sim10^9M_\odot$ the power exceeds $10^{-2}\LEdd$, indicating that $\Pradiation\ge\Poutflow$, i.e. we should see a nuclear point source. For $M_\mathrm{BH}\ge10^{10}M_\odot$, the power exceeds $10^{-3}\LEdd$. Here, $\Poutflow$  could dominate over \Pradiation, if the black hole is in a low accretion state. }
   \label{fig1}
\end{center}
\end{figure}

For highly-luminous cool core clusters (with $\Lx\ge10^{45}\ergps$), the central AGN must be injecting on average $10^{45}\ergps$ into the surrounding medium. We do not know the black hole (BH) mass for most of these AGN but we expect that they lie between $10^9$ and $10^{10}\Msun$, based on the few reliable measurements in the literature ({\rm e. g.} M87, \cite[Macchetto et al. 1997]{Macchetto et al. 1997}). Given that these objects are in an exceptional environment and are active continuously, it is not clear that the standard $M_\mathrm{BH}-\sigma$ or $M_\mathrm{BH}-M_\mathrm{K}$ relations are relevant for them. For $M_\mathrm{BH}\sim10^9M_\odot$, this means that these black holes must be operating at high enough Eddington rates that they should be radiatively efficient (see Fig. \ref{fig1}). We would therefore expect to see an X-ray point source. 

We present a sample of strong cool core, highly-luminous clusters, for which there is {\emph no} evidence of a nuclear X-ray point source in the $Chandra$ images. Using these images, we derive upper limits of the nuclear luminosities with the web interface of {\sc pimms} (\cite[Mukai 1993]{Mukai 1993}), which converts a count rate into an expected flux. We also investigate whether there is a hidden power law in the X-ray spectra, but find no such evidence in any of our objects. Finally, we calculate the energy that must be injected by the AGN in order to offset the cooling (\Loutflow) of the ICM, and compare it with the nuclear luminosity of the AGN. 

\section{Results}

\begin{figure*}
\begin{center}
 \includegraphics[width=5.5in]{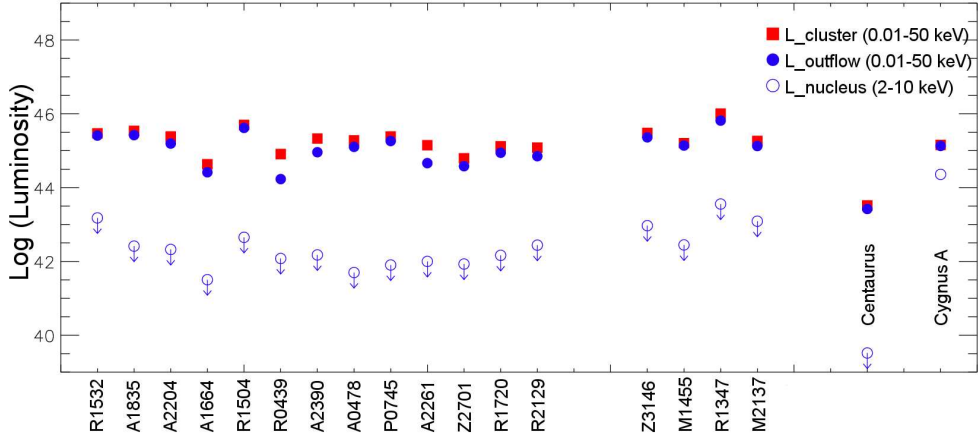} 
 \caption{Shown here, is our sample of 13 clusters with no clear X-ray point source. We also include 4 clusters with a possible point source (Z3246, M1455, R1347 and M2137); the Centaurus cluster, known to have a radiatively-inefficient AGN; and Cygnus A, a perfect counter-example with an obvious point source. For each object, we show the total X-ray luminosity (\Lcluster), the energy needed by the AGN to offset cooling (\Loutflow) and the derived upper limit for the nuclear X-ray luminosity (\Lnucleus). }
   \label{fig2}
\end{center}
\end{figure*}

\begin{figure*}
\begin{center}
 \includegraphics[width=3.7in]{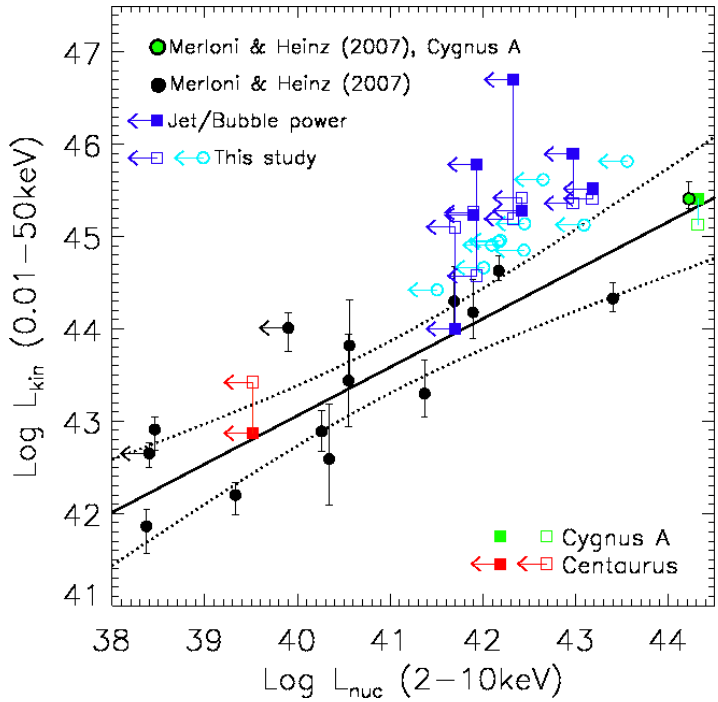} 
 \caption{Kinetic bolometric luminosity versus nuclear luminosity in the $2-10\keV$ band. The AGN of \cite[Merloni \& Heinz (2007)]{Merloni2007} are shown with black filled circles. These authors found that there was a statistical correlation between the kinetic bolometric luminosity and nuclear luminosity of an AGN. There sample included 15 AGN, and we plot with a black line, the linear fit, and with the dotted lines, the 2$\sigma$ limits. Our results are shown with the non-filled symbols. Clusters with estimates of their kinetic luminosity from measurements of the energy output from bubbles/jets are shown in squares and the estimate is shown with the filled square. }
   \label{fig3}
\end{center}
\end{figure*}

Our results are shown in Fig. \ref{fig2} and Fig. \ref{fig3}, and reveal a significant population of objects requiring high kinetic input from an AGN to offset cooling and/or high jet power, yet are without a detected X-ray nucleus. These objects appear to be radiatively inefficient with on average, $\Lkin/\Lnuc\ge200$. We examine 7 possible explanations as to why these objects appear to be so radiatively-inefficient.


\begin{itemize}
\item First, they could simply be strongly Doppler-suppressed. However, all jets would need to be aligned with the plane of the sky, and it is unlikely that all of our objects have jets with a preferred geometry.
\item Second, we could be dealing with Advection Dominated Accretion flows (\cite[Narayan \& McClintock 2008]{Narayan2008}), but it would be difficult to obtain jet powers of $10^{45}\ergps$.
\item Next, our objects could have magnetically-dominated black holes, but the mechanism responsible for creating jets in magnetically-dominated accretion discs still remains poorly understood, which makes it difficult to give any definitive conclusion.
\item They could also be highly-absorbed. The power must then emerge at longer wavelengths. However, only three of our central AGN have large IR luminosities. Furthermore, this luminosity is from $Spitzer$ data, and remains un-resolved for our objects. Most of the far IR emission is also likely due to young stars, since there is much evidence from excess blue light, that there is ongoing star formation in our objects (\cite[O'Dea et al. 2008]{O'Dea et al. 2008}).
\item We could be looking at the variability of how many X-rays are produced by the core (i.e. the radiative efficiency of the jets). Perhaps the radiative efficiency of a jet depends on how much interaction occurs between the jet and surrounding gas. Extremely powerful steady jets may create a channel which results in little such interaction. 
\item Next, they could be spin-powered black holes which could allow a strong jet to be created while keeping the AGN radiatively dim (see \cite[McNamara et al. 2009]{McNamara et al. 2009}). A high spin parameter would then be required, which according to our significant population of objects, would mean that highly rotating BHs are not rare.
\item Finally, a promising explanation could be that our objects have an ultra-massive black hole ($M_\mathrm{BH}\ge10^{10}M_\odot$), then the radiative power could fall well below the kinetic power as shown in Fig. \ref{fig1}. 
\end{itemize}

\section{Summary}

We have identified a population of objects which require powerful jets to be present but have no X-ray detectable nucleus. We have also identified a range of possible explanations, each of which carries significant implication for the origin and operation of jets. The black holes may be ultramassive ($M_\mathrm{BH}\gg10^9M_\odot$), or have very high spin, or be highly obscured. They may also be mostly off, yet unobservable when they are switched on, or have highly radiatively inefficient jets.

\end{document}